# Direct Observation of Cycloidal Spin Modulation and Field-induced Transition in Néel-type Skyrmion-hosting VOSe$_2$O$_5$

Takashi Kurumaji[1]*, Taro Nakajima[1†], Artem Feoktystov[2], Earl Babcock[2], Zahir Salhi[2], Victor Ukleev[1,3], Taka-hisa Arima[1,4], Kazuhisa Kakurai[1,5], and Yoshinori Tokura[1,6]

[1] *RIKEN Center for Emergent Matter Science (CEMS), Wako 351-0198, Japan,*

[2] *Forschungszentrum Jülich GmbH, Jülich Center for Neutron Science (JCNS) at Heinz Maier-Leibnitz Zentrum (MLZ), Garching 85748, Germany,*

[3] *Laboratory for Neutron Scattering and Imaging (LNS), Paul Scherrer Institute (PSI), CH-5232 Villigen, Switzerland,*

[4] *Department of Advanced Materials Science, The University of Tokyo, Kashiwa 277-8561, Japan,*

[5] *CROSS-Tokai, Research Center for Neutron Science and Technology, Tokai, Ibaraki 319-1106, Japan,*

[6] *Department of Applied Physics, The University of Tokyo, Tokyo 113-8656, Japan.*

*Current address: Department of Advanced Materials Science, the University of Tokyo, Kashiwa 277-8561, Japan.

†Current address: Institute for Solid State Physics, The University of Tokyo, Chiba 277-8581, Japan.

**Abstract**

**We investigate the spin rotational structure of magnetic skyrmions in a tetragonal polar magnet VOSe$_2$O$_5$ via polarized small-angle neutron scattering (SANS). Spin polarization analysis of the scattered neutrons provides consistent evidence for the cycloidal spin modulation in all the incommensurate phases at zero and non-zero magnetic field along the *c* axis, including the triangular skyrmion-lattice phase. In the vicinity of the skyrmion phase, we performed extensive SANS measurements to unravel a field-induced incommensurate phase (IC-2 state). We discuss the possibility of anisotropic double-*q* state as an alternative spin structure to provisional square skyrmion-lattice state.**

# 1. Introduction

Magnetic domain walls (DW), chiral soliton-lattices, and skyrmions are spatially varying noncollinear and/or noncoplanar spin arrangements in magnets characterized by nontrivial winding number in 1D or 2D space [1-3]. These spin structures, when isolated in ferromagnetic background, can be viewed as particles, and mobile via strong coupling with conduction electrons, hence being promising candidates for current-driven information carriers in spintronics devices [4,5]. These structures are characterized by spin helicity $\gamma$ which represents how the magnetic moments are twisted with respect to the spatial coordinates [1]. Spin axes rotate in the plane parallel to the propagation vector ($q$ vector) in a Néel-type DW/skyrmion with $\gamma = 0,$ and $\pi$ (Figs. 1(a, c)), which is in contrast with the Bloch-type spin configurations for $\gamma = \pm\frac{\pi}{2}$ (the signs define the handedness), where the spin rotates around the $q$ vector (Figs. 1(b, d)). Recent studies have revealed a significant role of $\gamma$ in the current-induced motion of DW/skyrmions through spin-orbit and spin-transfer torques [5-10]. These features provoke pressing interest in experimental techniques to distinguish between proper-screw and cycloidal spin configurations of these topological spin solitons.

Since the discovery of skyrmion-lattice (SkL) state in chiral magnets [11,12], recent experimental efforts have revealed that various classes of materials including bulk compounds [13] and thin-films/heterostructures host skyrmions with different helicities as well as vorticities [14]. The skyrmion types are mostly selected by Dzyaloshinskii-Moriya interaction originating from inversion-symmetry breaking [15,16]. Recent studies have discovered that these materials harbor versatile magnetic phase diagrams including both equilibrium and metastable skyrmions with different lattice forms such as triangular and square SkL states [17-21]. These rich features are generated from subtle balance among thermal fluctuations, Zeeman energy, magnetic anisotropy, and perhaps multi-spin exchange interactions [17,22-26]. Development of widely applicable methods to determine the spin configurations as well as the SkL form is essential for understanding the link among spin textures and their stability.

In order to determine the orientations of the magnetic moments composing skyrmions and to monitor the phase transitions in external fields, multiple experimental techniques are available, including Lorentz transmitting electron microscopy (LTEM) [12,27,28], spin polarized scanning tunneling microscopy (SPSTM) [17,29], spin-polarized low-energy electron microscopy (SPLEEM) [30], nitrogen-vacancy (NV) center based microscopy [31,32], and circular-dichroism in magnetic x-ray scattering [33,34]. Although LTEM has been well established to detect Bloch-type DW/skyrmions, it has only a limited sensitivity to Néel-type configuration due to geometrical condition for the electron-beam deflection by spins [27,28]. The other techniques have been proved effective to distinguish the types of skyrmions while they are mostly available at, or sensitive to the surfaces of specimens.

We here focus on polarized small-angle neutron scattering (PSANS) with longitudinal polarization analysis, which can distinguish between Bloch and Néel-type spin modulations and is broadly applicable to monitor the magnetic phase transitions in bulk compounds [35]. Up to now, Bloch-type magnetic modulations in magnetic fields were partly confirmed in the previous study on MnSi [36]. As for Néel-type skyrmions, a previous PSANS study in polar GaV$_4$S$_8$ [37] reported the evidence of cycloidal spin modulation in magnetic phases, while the intensity for skyrmions is hard to separate from that for the multi-domain state of the single-$q$ cycloidal phase. Direct observation of Néel-type skyrmions via PSANS has remained to be established.

In this manuscript, we report the PSANS observation of the triangular-lattice state of the Néel-type skyrmion in VOSe$_2$O$_5$. This material forms a square lattice of magnetic vanadium ions (V$^{4+}$: $S = 1/2$) (Fig. 1(e)) belonging to a tetragonal polar group (*P4cc*) [38-40], and has recently been identified to host the SkL state via unpolarized SANS [41]. Thermo-equilibrium phase diagram for the magnetic field (*H*) along the *c* axis is reproduced in Fig. 1(f), which contains incommensurate spin states for cycloid (IC-1), triangular SkL (*A*), and IC-2 states together with paramagnetic (PM) and field-induced ferrimagnetic (FM) states. The spin structure in IC-2 state was provisionally assigned to the square SkL state based on the theories [22,23] and the observed four-fold SANS pattern in terms of the in-plane magnetic modulation [41]. We performed further experiment to address this hypothesis, and have fond that an anisotropic double-$q$ [22,42] structure, instead of the square skyrmion lattice state, is more reasonable to explain the observed field evolution of SANS intensity.

## 2. Experimental

Polarized and unpolarized SANS measurements were performed at KWS-1 with the previously used assembled single crystals [41]. We introduce a laboratory coordinate system, *xyz*, as shown in the schematic experimental configuration (Fig. 1(g)). The sample was mounted in a closed-cycle $^4$He refrigerator, so that the crystallographic *a*, *b*, and *c* axes were parallel to the *x*, *y*, and *z* axes, respectively. The direction of incident neutron spins was controlled to be parallel or antiparallel to the *z* axis by a spin flipper, and maintained by guiding fields. An incident spin-polarized neutron beam was obtained by a supermirror polarizer. The wavelength of the incident neutron beam was tuned to 10 Å. We employed a polarized $^3$He spin filter to analyze the spin states of the scattered neutrons [43,44]. The $^3$He polarization can be reversed with adiabatic fast passage [45], which enables us to measure the four combinations of neutron polarization incident upon and scattered by the sample. The decay time of the used $^3$He cells was approximately 100 hours, which was monitored through the transmission of the direct neutron beam. Two different cells were used, J1 and Puck, which contain 5.2 and 5.0 bar cm of $^3$He, respectively [46,47]. The flipping ratio and transmission of the filter were at the maximum 20 and 15 % and the minimum 17 and 9 %, respectively. Imperfection of beam polarization was measured periodically and corrected following Refs. [48,49]. To extract the magnetic correlations below $T_\mathrm{C}$ = 7.5 K, the spectrum measured above $T_\mathrm{C}$, typically at 8 K, was chosen as background. The *H* on the sample was produced by a Helmholtz coil. Each polarized and unpolarized SANS intensity was taken by fixing the sample alignment with the *c* axis parallel to the incident neutron beam. We note that our observation is confined in the (00*L*) plane. Although it lacks a precise sensitivity to the spin correlation length along the $L$ direction, the polarization analysis provides precise knowledge on the direction of the spin-spiral plane.

Here, we describe the spatially-modulated magnetization with a wave vector $\boldsymbol{q}$ in the $ab$ plane as $\boldsymbol{m}(\boldsymbol{r}) = \mathrm{Re}[\boldsymbol{m}_0 e^{i\boldsymbol{q}\cdot\boldsymbol{r}}]$ ($\boldsymbol{m}_0$ is a complex vector, $\boldsymbol{r} = (r_a, r_b, r_c)$ is the position vector in terms of the spatial coordinate, $abc$, in a crystal). In SANS, magnetic Bragg reflections appear when the scattering vector $\boldsymbol{Q}$ ($= \boldsymbol{k}_\mathrm{f} - \boldsymbol{k}_\mathrm{i}$) coincides with $\pm\boldsymbol{q}$. Intensities of the magnetic reflections in the non-spin-flipping (NSF), and spin-flipping (SF) channels are given by (see Eqs. (32-33) in Ref. [50])

$$I_\mathrm{NSF} = I^{++} + I^{--} \propto 2|m_c^\perp(\boldsymbol{Q})|^2, \tag{1}$$

$$I_\mathrm{SF} = I^{+-} + I^{-+} \propto |m_a^\perp(\boldsymbol{Q}) + im_b^\perp(\boldsymbol{Q})|^2 + |m_a^\perp(\boldsymbol{Q}) - im_b^\perp(\boldsymbol{Q})|^2. \tag{2}$$

$I^{\alpha\beta}$ ($\alpha, \beta = +$ or $-$) denotes the observed intensity when the incident and scattered neutrons are in the spin states $\alpha$ and $\beta$, respectively. The signs ($\pm$) stand for the $\pm z$ direction for the neutron-spin polarization, respectively. $\boldsymbol{m}(\boldsymbol{Q})$ is the Fourier transformation of the $\boldsymbol{m}(\boldsymbol{r})$ and $\boldsymbol{m}^\perp(\boldsymbol{Q})$ is the projection on the plane perpendicular to the $\boldsymbol{Q}$, i.e., $\boldsymbol{m}^\perp(\boldsymbol{Q}) = \boldsymbol{m}(\boldsymbol{Q}) - [\boldsymbol{m}(\boldsymbol{Q}) \cdot \hat{\boldsymbol{Q}}]\hat{\boldsymbol{Q}}$, where $\hat{\boldsymbol{Q}}$ is the unit vector along the $\boldsymbol{Q}$. The expressions for the cycloidal and proper-screw spin configurations propagating along the $a$ axis (Figs. 1(a,b)) are, respectively, given by

$$\boldsymbol{m}_\mathrm{cyc}(\boldsymbol{r}) = m_c \cos(Qr_a)\hat{\boldsymbol{r}}_c + m_a \sin(Qr_a)\hat{\boldsymbol{r}}_a, \tag{3}$$

$$\boldsymbol{m}_{\text{screw}}(\boldsymbol{r}) = m_c \cos(Qr_a)\,\hat{\boldsymbol{r}}_c + m_b \sin(Qr_a)\,\hat{\boldsymbol{r}}_b. \tag{4}$$

By substituting each $\boldsymbol{m}^{\perp}_{\text{cyc/screw}}(\boldsymbol{Q})$ for Eqs. (3)-(4) into Eqs. (1) and (2), one can find that the $I_{\text{SF}}$ is absent in the cycloidal but present in the proper-screw case. NSF scattering is produced by the $m_c$ component in both cases. Therefore, the measurement of $I_{\text{SF}}$ is applicable to rule out the proper-screw nature of the spin structure. This conclusion can be readily generalized for distinguishing between Néel and Bloch-type SkL states with a triple-$q$ vector in the *ab* plane.

## 3. Results and Discussion

Figures 2(a) and 2(b) show the observed NSF and SF scatterings, respectively, for the triangular SkL state in the *A* phase. Presence of intensity in NSF channel (Fig. 2(a)) and absence of that in SF channel (Fig. 2(b)) are consistent with the above anticipation for the Néel-type skyrmion state [51]. We also confirmed that the NSF channel reproduces the twelve-fold intensity peaks at $|Q| = 0.048$ nm$^{-1}$. As identified in the previous unpolarized SANS study [41], this pattern indicates the coexistence of two domains of triangular SkL state with one of the $q$ vectors locked along either *a* or *b* axis (Fig. 2(c)) together with the IC-1 or IC-2 states with $q\|a$ and $q\|b$. The emergence of the twelve-fold peak exclusively in the NSF channel enables us to verify the presence of a triangular-lattice state of skyrmions in the present sample along with the Néel-type nature of these skyrmions. As for GaV$_4$S$_8$ [37], the SANS intensity for the skyrmion phase forms a ring-like pattern: the $q$ vectors isotropically distribute in the rhombohedral (111) plane. As a result, it was difficult to tell the presence of the skyrmions region apart from random $q$-domain of cycloidal state by the PSANS study alone.

We performed a further analysis of the polarized small-angle neutron scattering (PSANS) intensity profile for the NSF channel (Fig. 3(a)) to resolve the twelve-fold spot pattern, as shown in Fig. 3(b). We integrated the PSANS intensity along the radial axis within $0.004 < |Q| < 0.0058$ Å$^{-1}$ (Fig. 3(a)) to obtain the azimuthal-angle ($\phi$) dependence of the intensities of magnetic Bragg scattering in the NSF channel. The result is shown in Fig. 3(c). We used a multi-Gaussian function:

$$I(\phi) = \sum_{n=0}^{5} I_{an} \exp[-(\phi - 60n - \phi_{0a})^2/2\Delta\phi_a^2] + \sum_{m=0}^{5} I_{bm} \exp[-(\phi - 30 - 60m - \phi_{0b})^2/2\Delta\phi_b^2], \quad (5)$$

for the fitting, where $I_{an}$ and $I_{bm}$ ($n, m$ = 1-5) are peak intensities for each Bragg peak, $\phi_{0a}$ and $\phi_{0b}$ are origin shifts, and $\Delta\phi_a$ and $\Delta\phi_b$ are widths of Bragg peaks belonging to $q_a$ and $q_b$-SkL domain, respectively. In Fig. 3(c), we decompose the fitting curve for each Bragg peak as shown by red and blue curves. As shown in Fig. 3(b), individual Bragg peaks can be identified within the standard deviation at each peak position. We note that the intensity variation of Bragg peaks for the $q_a$-SkL domain (red curves in Fig. 3(c)) appears stronger than that of the $q_b$-SkL domain (blue curves). In particular, intensities at $\phi \sim 120°$ and $300°$, diagonal with each other with respect to the detector center, are relatively weaker than the others. This may be due to a slight misalignment of the $c$ axis towards $\phi$ = 120° (or 300°). We also plot the $\phi$ dependence of the PSANS profile for the SF channel in Fig. 3(d), where no peak is observed.

We also distinguish between proper-screw and cycloidal spin configurations in the IC-1, and IC-2 state. Figure 2(d) shows the intensity pattern for the NSF channel in the IC-1 state, reproducing that in the unpolarized SANS study [41], i.e., four-fold peaks for $q\|a$ and $q\|b$, which are locked by tetragonal anisotropy. We also discerned that weak ring-like intensity is superimposed due to diffusive $q$-domains with random $q$ directions. Absence of the intensity in the SF channel (Fig. 2(e)) supports the cycloidal spin configuration in the IC-1 state. As assigned in Ref. [41], these results indicate that two kinds of single-$q$ domain with $q_a$ and $q_b$ dominates as shown in Fig. 2(f). We also proved the cycloidal spin configuration in the IC-2 state by the presence and absence of the PSANS intensity for NSF (Fig. 2(g)) and SF channel (Fig. 2(h)), respectively. The IC-2 state was previously proposed as the square-lattice of the Néel-type skyrmions. This hypothetical single-domain spin texture is expressed by the equivalent superposition of two cycloidal spin modulations propagating along $a$ and $b$ axes. We did not confirm whether the magnitude of these two modulations are equivalent or contain some sort of anisotropy because the possible multidomain state of the latter would be indistinguishable from the former. In the latter case, the spin texture is not the square SkL state, but is rather termed an anisotropic double-$q$ state proposed in theoretical works [22,42]. Schematic multidomain configuration is depicted in Fig. 2(i). Diffusive $q$-domains as observed by the ring-like pattern in Fig. 2(g) are not shown here.

In order to investigate the multiple-$q$ nature for the IC-2 phase, we monitor the field dependence of the unpolarized SANS intensity through the phase transition from IC-1 state to IC-2 state at $T = 6.67$ K for $H||c$. Prior to the measurement, we rotated the Helmholtz coil in the horizontal ($zx$) plane to apply $H$ along the $x$ ($a$) axis. The single domain of the IC-1 state with $q||b$ (Fig. 4(a)) was selected due to the anisotropy of the Zeeman energy for the cycloidal spin configuration [41]. After eliminating the $H$, we rotated the coil back to $H||z(c)$ configuration. We subsequently applied the field along the $c$ axis to take the minor loop as shown in Figs. 4(a)-(h). We find that even at $\mu_0 H = 5$ and 6 mT (Figs. 4(c-d)), in the center of the IC-2 phase (Fig. 1(f)), the SANS intensity for $q||a$ remains considerably weaker than that for $q||b$.

For more quantitative perspective, we plotted the field dependence of the averaged magnetic modulation wavelength $\lambda$ ($= 2\pi/q$) (Fig. 5(a)) and the intensities on the $ab$ plane for the respective areas parallel to $a$ and $b$ axes (Fig. 5(b)). As increasing the field, $\lambda$ increases and shows a kink at around 2.5 mT signaling the transition into the IC-2 state. We find that $\lambda$ shows a hysteresis in the backward scan for the minor loop (open symbols in Fig. 5(a)) only in the IC-2 region. This reflects the different spin textures in IC-1 and IC-2 state separated by a first-order phase boundary. Furthermore, the intensity for the horizontal direction ($q||a$) appears in the IC-2 state as shown in Fig. 5(b). Although we have no precise estimation of the peak width along the $c^*$ direction for each peak in this experiment as stated above, we note that the spin structure in the IC-2 state is highly anisotropic in the *ab* plane on the assumption of the similar extent of the spin correlation lengths along the $c^*$ direction for both $q\|a$ and $q\|b$ components.

Here, we consider another possibility for the spin texture in the IC-2 phase: an anisotropic double-$q$ state. This was initially proposed as a stable spin texture in a chiral magnet with tetragonal single-site anisotropy and/or compass-type anisotropic exchange interaction [22,42]. In the present noncentrosymmetric magnet of $V^{4+}$ ions with $S = 1/2$, compass-type anisotropic exchange interactions are allowed through the spin-orbit interaction [52-54]. The effective tetragonal anisotropy is confirmed from the fact that the $q$ vectors favor to align along the *a* or *b* axis [41]. The modulated component of the spin density $\boldsymbol{S}(r_a, r_b)$ is expressed as

$$\boldsymbol{S}(r_a, r_b) = C_a(\sin qr_a, 0, \cos qr_a) + C_b(0, \cos qr_b, \sin qr_b), \quad (6)$$

where the amplitudes for the cycloidal modulations along *a* and *b* axes are different with each other ($C_a \neq C_b$). As shown in Fig. 2(i), inequivalent amplitudes break the tetragonal symmetry, and global spin texture does not hold topological number.

On the basis of this spin structure model, we can describe the evolution of the intensity in the minor loop as schematically illustrated in Fig. 5(c). Starting from the purified single-domain of the IC-1 state (1) with $q_b$ in zero field, the intensity for the minor $q_a$ modulation develops with $H\|c$ (2) due to the formation of the IC-2 state with the primal $q_b$. Further increase of the field induces the FM state (3) at some parts of the pieces of crystals erasing the memory of the $q$ orientation. In the backward scan, the region for the FM state returns to the IC-2 state (4) with randomly selecting the primal $q$ vectors, which is compatible with the observation where the intensities for $q\|a$ and $q\|b$ evolve parallel with each other (open symbols in Fig. 5(b)). Inequality of these intensities can be ascribed to the remnant domain for the IC-2 state, which does not lose the memory from the initial single domain state. Eventually, the system comes back to the IC-1 state (5) in zero field with contamination of the domain with $q_a$, as observed in Fig. 4(h).

The anisotropic double-$q$ state is an intermediate state between the square SkL and the single-$q$ state. In previous studies, square SkL states have been exclusively observed at low temperatures as either the metastable [19,20] or the thermo-equilibrium states [21]. Thus, we expect that the square SkL state in $VOSe_2O_5$ would position, if present, at relatively low temperatures as well. Our experiment, however, suggests that such a spin modulation is unfavored by easy-plane anisotropy below 4 K [41]; around 6 K we anticipate that thermal fluctuations suppress the formation of the square SkL state and rather push the stability to the side of the single-$q$ state, and as a result favor the anisotropic double-$q$ state. Interestingly, thermal fluctuations also play an essential role for the magnetism in trigonal $GaV_4S_8$ [37,55], corresponding magnetic phase is lacked. Since the theories [22, 42] were performed at the absolute zero temperature, further theoretical investigation applicable at finite temperatures would be desirable to clarify the effect of the thermal fluctuations in tetragonal systems. Lastly, we leave a note that we cannot completely rule out the multidomain formation of the single-$q$ cycloidal state together with a small portion of the square skyrmion-lattice state in the IC-2 region. However, the current observation is hard to explain if the square SkL state is dominantly formed in the IC-2 phase. Future research via LTEM/MFM on $VOSe_2O_5$ is expected to provide complementary real-space information of the IC-2 state to prove the proposed spin structure in this work.

## 4. Conclusions

In conclusion, we directly determined the orientation of the magnetic moments composing of the SkL state via SANS measurement with a polarization analysis to confirm the Néel-type skyrmions in the tetragonal polar magnet $VOSe_2O_5$. We also obtained the evidence that the spin configuration for the other modulated magnetic phases surrounding the SkL phase, i.e., the IC-1 and IC-2 states, are both described with the cycloidal modulations. On the basis of the field evolution of the SANS pattern from the IC-1 (single-$q$ cycloid) to the IC-2 states, we propose the anisotropic double-$q$ state for the IC-2 phase. The present study provides a useful guiding principle to apply this procedure to determination of complex or multiple helical modulations anticipated to occur in skyrmion-hosting materials.

**Acknowledgments**

This research was supported in part by Grant-In-Aid for Young Scientists(B) No. 17K14351 from the Japan Society for the Promotion of Science (JSPS), and CREST No. JPMJCR1874 from the Japan Science and Technology Agency (JST). One of the authors (KK) acknowledges the support by the Helmholtz International Fellow Award.

# Figure Captions

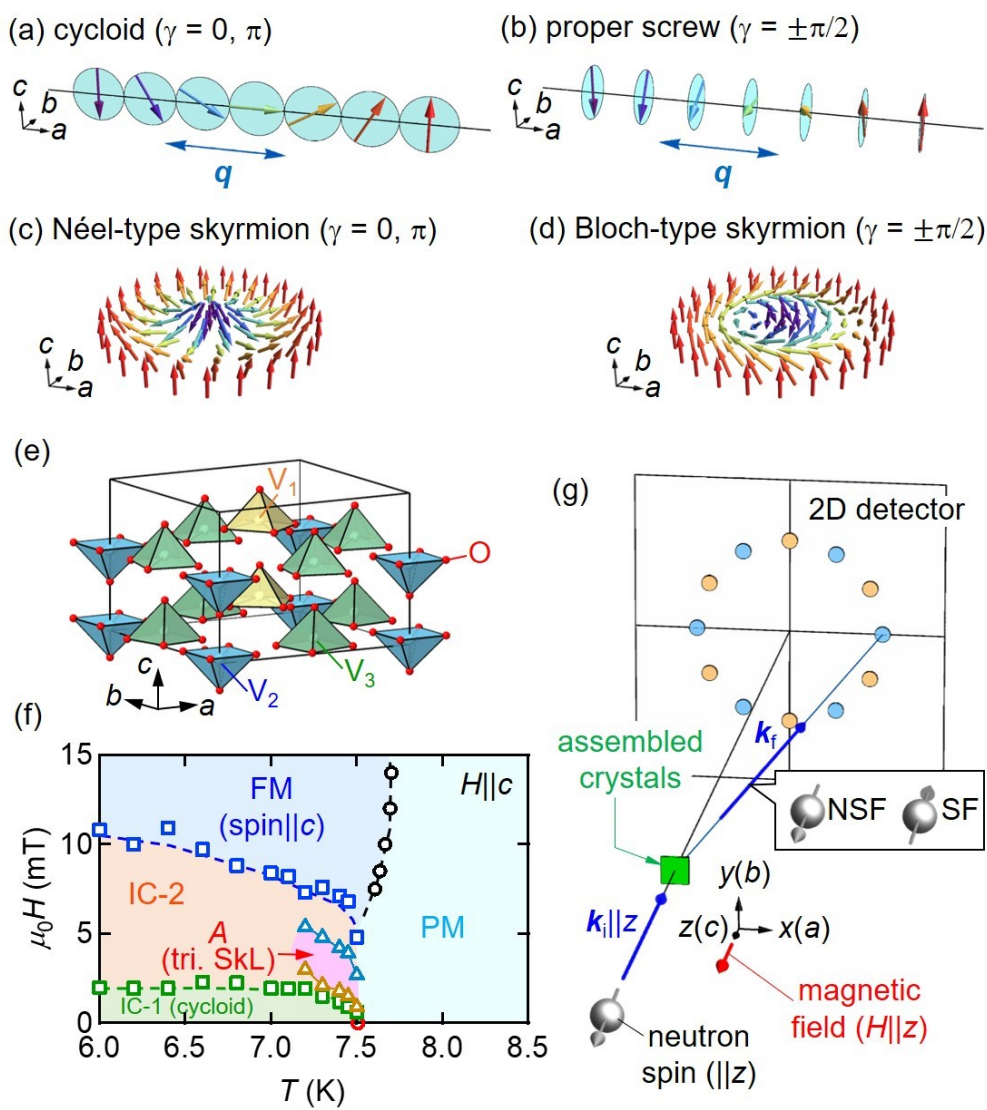

**Fig. 1** (Color online) (a) [(b)] Schematic cycloidal [proper-screw] spin configuration with the $q$ vector in the $ab$ plane. The type of helicity $\gamma$ is linked by the spin rotation direction with respect to the $q$ vector. (c) [(d)] Spin configuration in a Néel- [Bloch-]type skyrmion. (e) Crystal structure of $VOSe_2O_5$. Yellow, blue, and green tetragonal pyramids highlight respective $VO_5$ polyhedra. $Se_2O_5^{2-}$ molecular anions are not shown. (f) $H$-$T$ phase diagram

for assembled coaligned single crystals in $H\|c$, reproduced from Ref. [41]. (g) Schematic PSANS setup used in the present study. The case for the incident neutron spin (gray arrow) polarized in the $+z$ direction is depicted. The assembled crystals (green rectangle) are oriented with the $c$ axis parallel to the $z$ axis. Blue arrows indicate the incident neutron beam direction ($\boldsymbol{k}_\mathrm{i}$) parallel to the $z$ axis and one of the final beam directions ($\boldsymbol{k}_\mathrm{f}$) scattered by a spin modulation. The $H$ (red arrow) is applied along the $z$ axis except for the field-trained process for the experiment focusing on the IC-2 state (see text). Blue and orange spots on the 2D detector indicate the doubling of six-fold magnetic Bragg peaks, for the domains of triangular SkL state with one of the triple-$q$ vectors oriented along $a$ and $b$ axes, respectively.

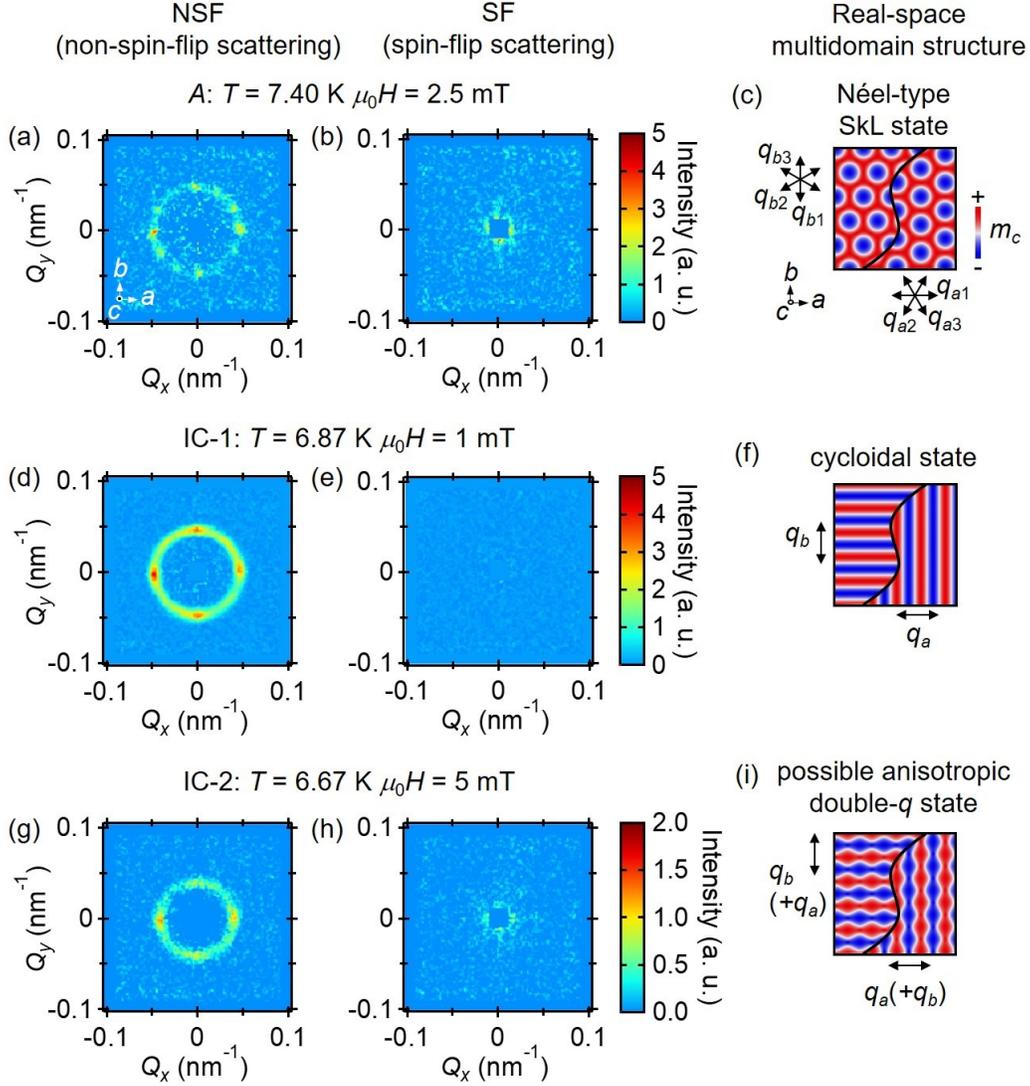

**Fig. 2** (Color online) (a), (d), (g) PSANS patterns of NSF and (b), (e), (h) SF channel at each magnetic phase. (a)-(b): *A* phase. (d)-(e): IC-1 phase. (g)-(h): IC-2 phase. (c), (f), (i) Color plot of $m_c$ component for the schematic multidomain state of each magnetic phase. Black arrows indicate the crystal axes or principal *q*-vectors for each domain. In (c), possibly coexisting IC-1 and/or IC-2 states are not shown. In (i), the spin state is described by Eq. (6), where a uniform magnetization component is omitted.

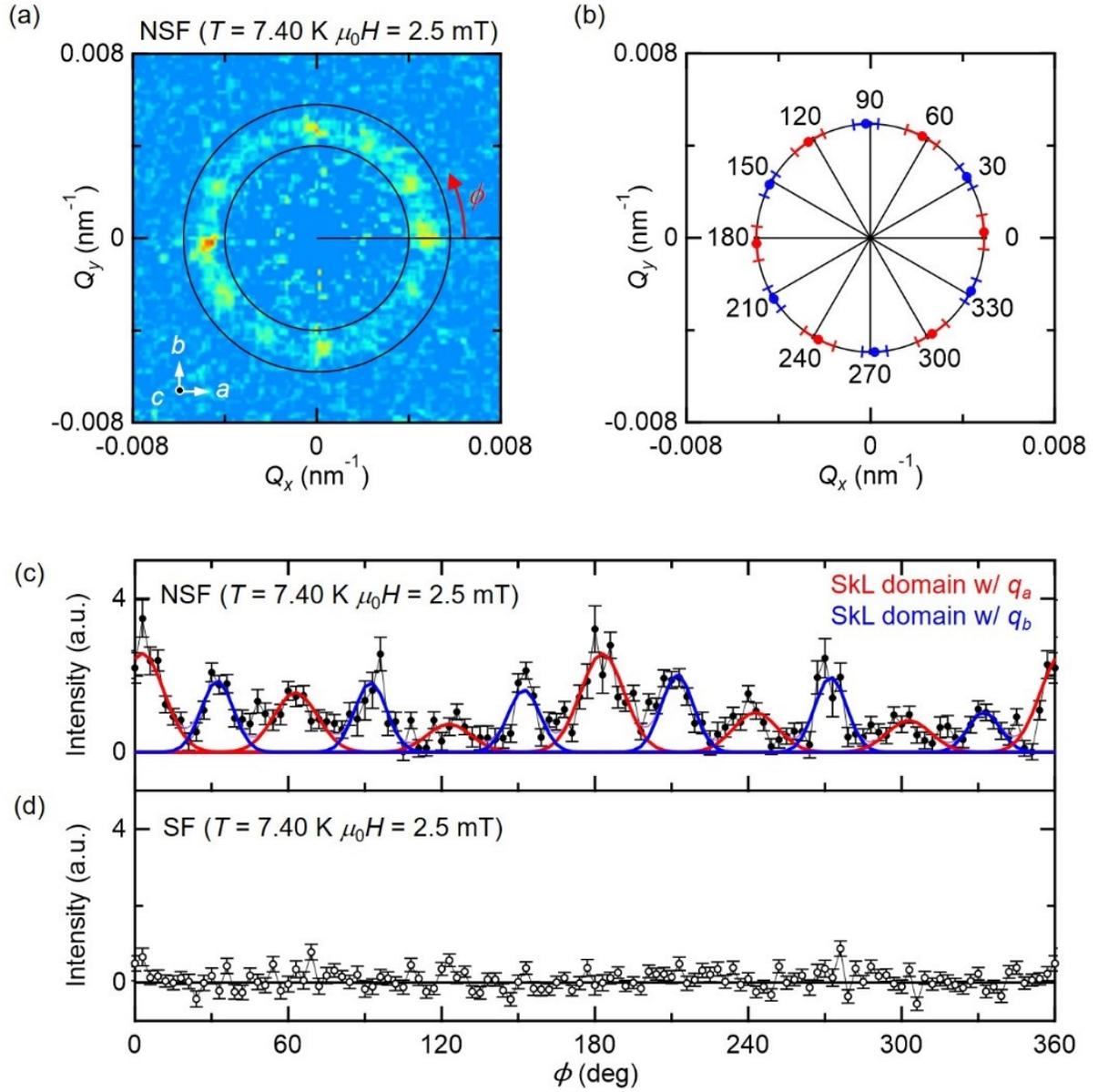

**Fig. 3** (Color online) (a) PSANS pattern of the NSF channel for the SkL state in the *A* phase measured at $T = 7.40$ K in the magnetic field of $\mu_0 H = 2.5$ mT along the *c* axis (reproduced from Fig. 2(a)). Solid circles define the boundaries for the region $0.004 < |Q| < 0.0058$ Å$^{-1}$ where SANS intensity was integrated to produce the azimuthal angle ($\phi$) plot shown in (c). (b) Red (blue) circles indicate the positions of the six-fold magnetic Bragg peaks for the domain of the SkL with one of the $q$ vectors, $q_a$ ($q_b$), along the $a$ ($b$) axis. Peak positions as well as the error bars are the center and the standard deviation of the each gaussian function obtained by fitting the data in (c) with Eq. (5). The definition of $\phi$ is also displayed. (c) and (d) $\phi$ dependence of the intensity of the PSANS pattern integrated along the radial axis for (c) NSF and (d) SF channels, respectively, obtained from the data presented in Figs. 2(a,b). Error bars show standard deviations. Red and blue curves in (c) are the multi-Gaussian fits for magnetic Bragg peaks belonging to $q_a$ and $q_b$-SkL domains, respectively. Individual Gaussian curves are decomposed from the resultant fitting curve for Eq. (5).

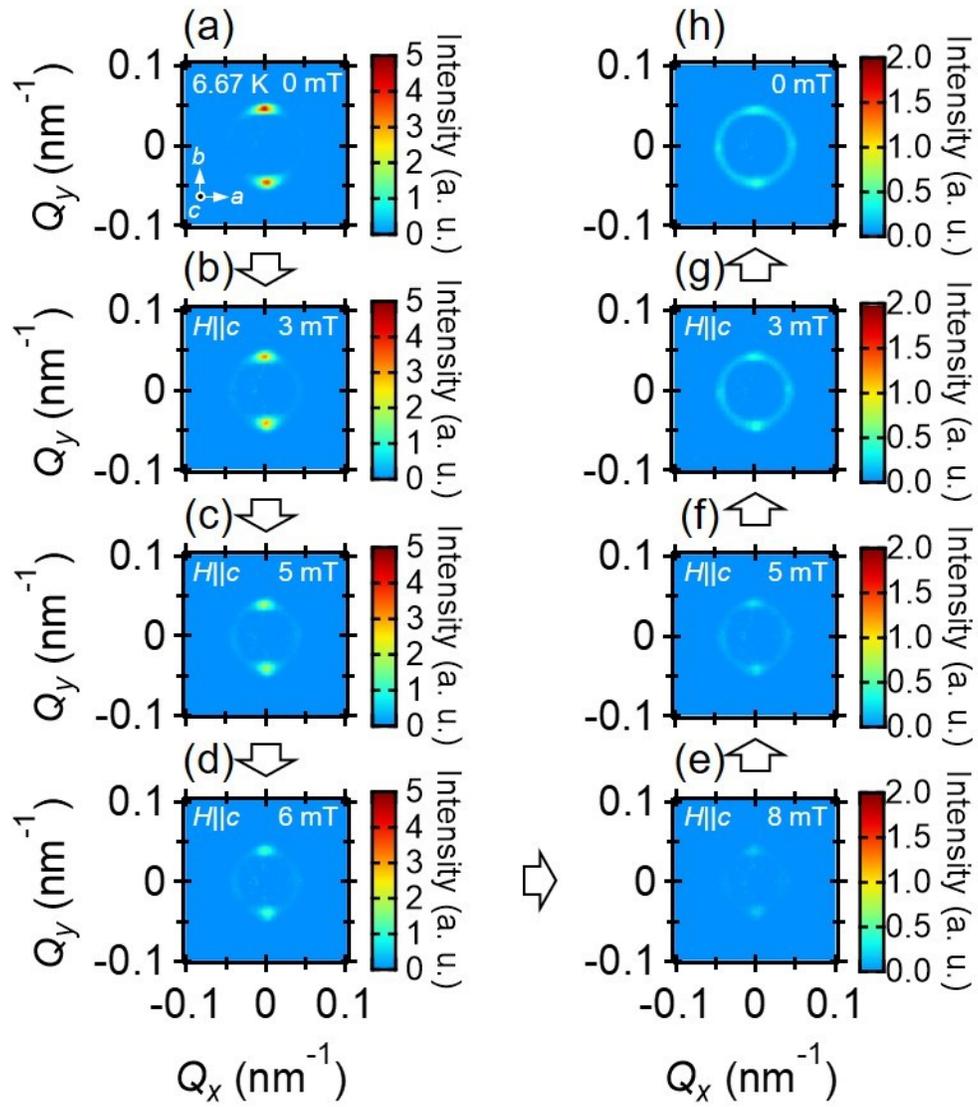

**Fig. 4** (Color online) (a)-(h) Variation of SANS patterns at 6.67 K with $H\|c$. The $H$ changes from (a) 0 mT to (e) 8 mT and back to (h) 0 mT in the sequential order indicated by white arrows.

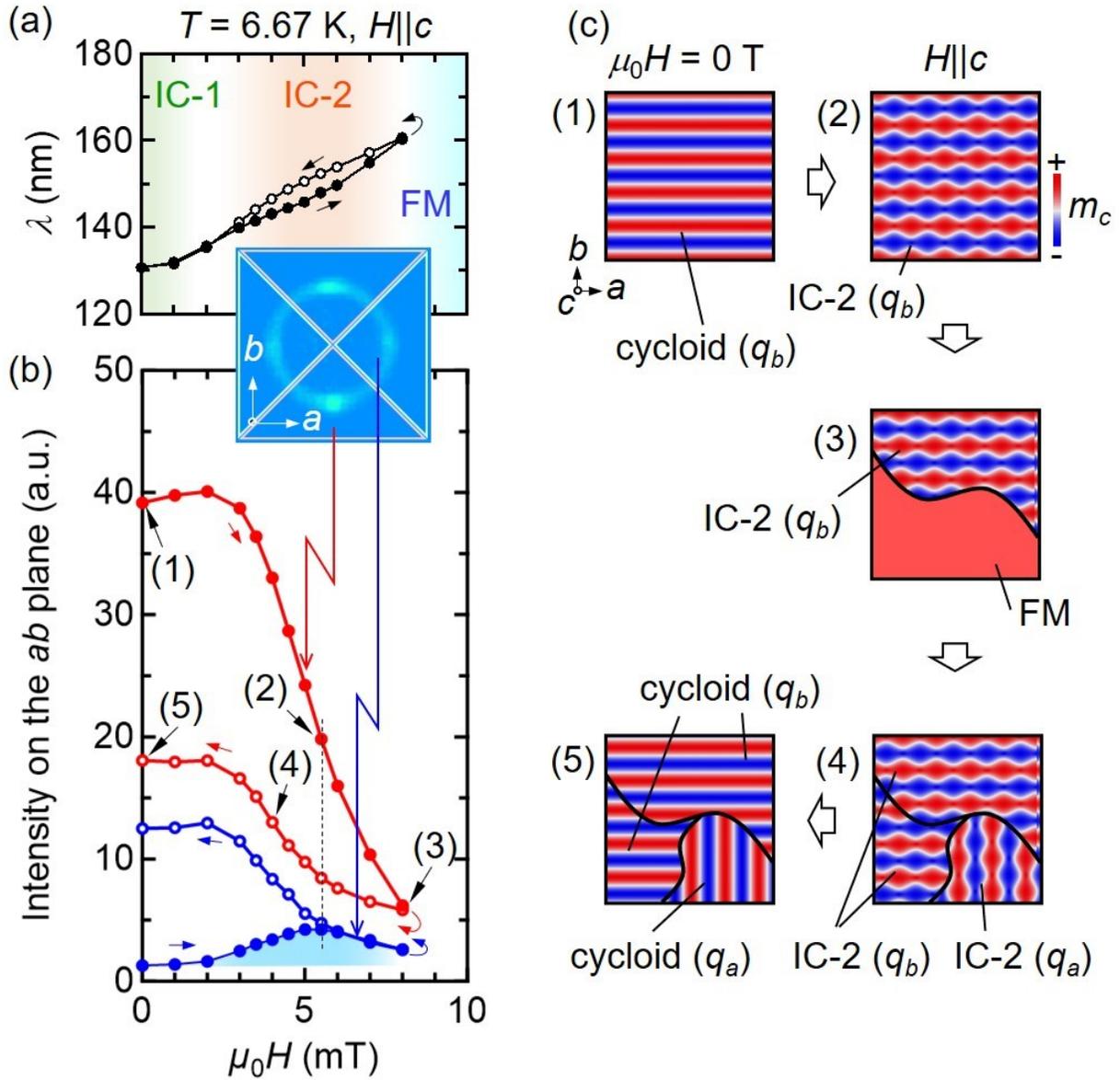

**Fig. 5** (Color online) (a)-(b) Magnetic-field dependence of (a) magnetic modulation wavelength ($\lambda$) and (b) SANS intensities on the $ab$ plane at 6.67 K in $H\|c$. Red (blue) circles are intensities for the regions parallel to $a$ ($b$) axes as indicated in the inset. Closed (open) circles are for the $H$-increasing (decreasing) scan. (c) Schematic magnetic domain structure in the field-scan process along the minor loop from (1) to (5) as indicated in (b).